\def\buildextended{1} % arXiv: extended version (full proofs and appendices)
\documentclass[conference]{IEEEtran}
\usepackage{cite}
\usepackage{amsmath,amssymb,amsfonts}
\usepackage{wasysym}
\usepackage{amsthm}
\usepackage{mathtools}
\usepackage{graphicx}
\usepackage{textcomp}
\usepackage{placeins}
\usepackage{booktabs}
\usepackage[dvipsnames,table]{xcolor}

\usepackage{url}
\usepackage{tikz}
\usetikzlibrary{positioning,calc,shapes.geometric,shapes.misc,fit,backgrounds,decorations.pathreplacing}
\usepackage{pgfplots}
\pgfplotsset{compat=1.16}
\usepackage[most]{tcolorbox}

\usepackage[hidelinks]{hyperref}
\hypersetup{pdfcreator={},pdfproducer={}}
\pdftrailerid{}
\newcommand{\artifacturl}{\url{https://github.com/anmolpandey299/al-synth}}
\providecommand{\buildextended}{0}
\newcommand{\scalingheight}{5.0cm} % Fig. 5 height (tuned to fill the last page)
\newif\ifextended \ifnum\buildextended=1 \extendedtrue\fi

\definecolor{ink}{HTML}{2A3A45}     % dark slate: strokes + text
\definecolor{coral}{HTML}{D26A52}    % warm accent: violation / exponential / emphasis
\definecolor{teal}{HTML}{3F8A78}     % cool accent: safe / solver / secondary
\definecolor{gold}{HTML}{D6A43A}
\definecolor{mist}{HTML}{EEF2F5}     % pale blue-gray fill
\definecolor{paleC}{HTML}{F4E6E0}    % pale coral fill
\definecolor{paleG}{HTML}{E7EFEA}    % pale teal fill (key-result boxes)
\definecolor{danger}{HTML}{B23B2E}
\definecolor{ruleg}{HTML}{C3CDD3}    % light rule gray
\tikzset{>=stealth, line cap=round, line join=round}

\tikzset{
  chk/.pic={\draw[teal,line width=1.0pt] (-0.075,0.01) -- (-0.02,-0.06) -- (0.085,0.07);},
  crs/.pic={\draw[coral,line width=1.0pt] (-0.06,0.06) -- (0.06,-0.06);
            \draw[coral,line width=1.0pt] (-0.06,-0.06) -- (0.06,0.06);}
}

\newtcolorbox{keyresult}{enhanced, breakable=false, colback=paleG, colframe=teal,
  boxrule=0.4pt, leftrule=2.4pt, arc=2pt, left=7pt, right=6pt, top=4pt, bottom=4pt,
  before skip=6pt, after skip=6pt}

\newcommand{\Low}{\mathsf{low}}
\newcommand{\Hi}{\mathsf{hi}}
\newcommand{\Obs}{\mathsf{obs}}
\newcommand{\Ctwo}{C_2^{D}}
\newcommand{\Crel}{C_2^{\mathrm{rel}}}

\newcommand{\Y}{\textcolor{teal}{$\bullet$}}
\newcommand{\N}{\textcolor{ruleg}{$\circ$}}
\newcommand{\Yr}{\textcolor{coral}{$\bullet$}}
\newcommand{\Yc}{\textcolor{gold}{\LEFTcircle}}

\theoremstyle{plain}
\newtheorem{theorem}{Theorem}
\newtheorem{lemma}{Lemma}
\newtheorem{corollary}{Corollary}
\theoremstyle{definition}
\newtheorem{definition}{Definition}

\def\BibTeX{{\rm B\kern-.05em{\sc i\kern-.025em b}\kern-.08em
    T\kern-.1667em\lower.7ex\hbox{E}\kern-.125emX}}

\begin{document}
\bstctlcite{IEEEexample:BSTcontrol}

%------------------------------------------------------------------------------
\title{Local Allow, Global Deny:\\
Why Per-Action Authorization Does Not Compose\\ in LLM Agents}

% Funding, if any, goes here per the template (unnumbered first-page footnote):
% \thanks{This work was supported by ...}  (also uncomment \IEEEoverridecommandlockouts)

% Authors, order, and affiliations exactly as accepted (EasyChair submission 9477).
% Optional first line per template: \textit{dept. name of organization}\\
\author{%
\IEEEauthorblockN{Anmol Pandey}
\IEEEauthorblockA{\textit{University of Warwick}\\
Coventry, United Kingdom\\
Anmol.Pandey@warwick.ac.uk}
\and
\IEEEauthorblockN{Aditya Jain}
\IEEEauthorblockA{\textit{University of Warwick}\\
Coventry, United Kingdom\\
aditya.jain.5@warwick.ac.uk}
\and
\IEEEauthorblockN{Liang Chen}
\IEEEauthorblockA{\textit{University of Hertfordshire}\\
Hatfield, United Kingdom\\
l.chen26@herts.ac.uk}
\and
\IEEEauthorblockN{Carsten Maple}
\IEEEauthorblockA{\textit{University of Warwick}\\
Coventry, United Kingdom\\
cm@warwick.ac.uk}}

\maketitle
% IEEE copyright notice required for the accepted manuscript (drawn in the top margin; no reflow)
\begin{tikzpicture}[remember picture,overlay]
\node[anchor=north,yshift=-0.22in,text width=0.9\paperwidth,align=center,font=\footnotesize] at (current page.north)
 {\copyright\,2026 IEEE. Personal use of this material is permitted. Permission from IEEE must be obtained for all other uses, in any current or future media, including reprinting/republishing this material for advertising or promotional purposes, creating new collective works, for resale or redistribution to servers or lists, or reuse of any copyrighted component of this work in other works. Accepted for publication in the proceedings of IEEE TPS 2026.};
\end{tikzpicture}

%------------------------------------------------------------------------------
\begin{abstract}
Autonomous LLM agents act through tools under \emph{per-action} authorization: a
local guard screens each tool call in isolation. We show that local authorization
does not compose into global authorization. A \emph{compositional authorization
failure} (CAF) is a locally authorized execution, or a pair of locally authorized
executions, that violates a global policy spanning intent, purpose, provenance,
consent, and delegation. We characterize CAFs along two axes, expressiveness
(single-trace safety versus relational $k$-safety) and observability (a ladder of
monitor profiles). Our central result is an impossibility: relational purpose
violations form a $2$-safety hyperproperty with no per-trace locus of blame, so a
per-action monitor that never raises a false alarm detects none of them, independent of
the underlying model or decision procedure; they become visible only under
cross-execution monitoring, which none of the six guardrails we survey implements. We
make this constructive with \textsc{AL-Synth}, a SAT/SMT synthesizer (using no language
model) that emits minimal CAF witnesses, each checked against an exhaustive oracle. The
empirical signature is sharp: across nine LLM judges the per-action view approves the
synthesized witness almost without exception, blocking it at a rate comparable to a
benign control, yet the same models flag it when shown the two executions together.
Per-action authorization is therefore structurally insufficient for relational policy
conjuncts not reducible to single-trace safety: to bound them, a monitor must compare
executions or accept false alarms.
\end{abstract}


%==============================================================================
\section{Introduction}
%==============================================================================
Agents built on large language models act through tools: they read mail, write files,
call web services, and delegate subtasks. The dominant safeguard is \emph{per-action}
authorization. A guard inspects each tool call in isolation and decides whether it is
permitted, and a sequence of approved steps is treated as an approved sequence.

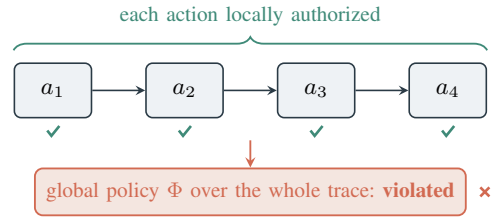
\begin{figure}[t]
\centering
\begin{tikzpicture}[font=\small]
  \tikzset{chip/.style={draw=ink,line width=0.6pt,rounded corners=2.5pt,fill=mist,
           minimum width=1.02cm,minimum height=0.72cm,inner sep=1pt}}
  \node[chip] (a1) {$a_1$};
  \node[chip,right=0.7cm of a1] (a2) {$a_2$};
  \node[chip,right=0.7cm of a2] (a3) {$a_3$};
  \node[chip,right=0.7cm of a3] (a4) {$a_4$};
  \foreach \a/\b in {a1/a2,a2/a3,a3/a4}{\draw[->,ink,line width=0.6pt] (\a)--(\b);}
  % per-step local approval: a teal check under each action (green at every step)
  \foreach \n in {a1,a2,a3,a4}{\path ($(\n.south)+(0,-0.17)$) pic{chk};}
  % top: teal brace = every local check passes
  \coordinate (tl) at ($(a1.north west)+(0,0.16)$);
  \coordinate (tr) at ($(a4.north east)+(0,0.16)$);
  \draw[decorate,decoration={brace,amplitude=5pt},teal,line width=0.7pt] (tl) -- (tr);
  \node[teal,font=\footnotesize,anchor=south] at ($(tl)!0.5!(tr)+(0,0.2)$)
       {each action locally authorized};
  % bottom: coral pill = global policy violated (red over the whole)
  \coordinate (mid) at ($(a1)!0.5!(a4)$);
  \node[draw=coral,fill=paleC,rounded corners=3pt,line width=0.7pt,below=1.02cm of mid,
        minimum width=5.7cm,minimum height=0.62cm,inner sep=3pt,text=coral,font=\footnotesize] (phi)
       {global policy $\Phi$ over the whole trace: \textbf{violated}};
  \path (phi.east) ++(0.24,0.0) pic{crs};
  \draw[->,coral,line width=0.7pt] ($(mid)+(0,-0.66)$) -- (phi.north);
\end{tikzpicture}
\caption{A compositional authorization failure, single-trace case. Every action passes
its local check (top), yet the composed trace violates the global policy $\Phi$
(bottom), with no single offending step. The relational purpose failure is witnessed
instead by a pair of individually compliant executions (Figure~\ref{fig:rel}).}
\label{fig:caf}
\end{figure}

This trust model is wrong in a way that matters: \emph{local authorization does not
compose}. A sequence of individually permitted actions can violate a global policy,
alone or together with a second, equally compliant execution. We call such a witness a
\emph{compositional authorization failure} (CAF), illustrated in Figure~\ref{fig:caf}.

This is the agent-era form of a classical fact, that security properties need not compose~\cite{mccullough1988noninterference}, sharpened by two features. First, some
policies that agents must respect are
\emph{hyperproperties}~\cite{clarkson2010hyperproperties}: they constrain sets of executions, not single runs. A purpose constraint says that what an agent discloses must not depend on
context its declared purpose does not authorize, and checking this requires comparing
runs. Second, the guards being deployed are single-trace execution monitors in
Schneider's sense~\cite{schneider2000enforceable}: each observes one run and decides.
The mismatch is a gap in expressive power, not an engineering gap that a better
per-trace monitor could close.

Our results are negative and constructive. We prove that one class of failure, the
relational purpose violation, has \emph{no per-trace locus of blame}: a per-trace
monitor that never flags a compliant trace detects \emph{none} of these violations,
because each trace alone is compliant and the violation lives only in the relationship
between runs. It becomes detectable only by comparing executions, which none of the six
guardrails we survey does. We make the result mechanical with \textsc{AL-Synth}, a
bounded model checker built on a SAT/SMT solver, with no language model, that
synthesizes minimal CAF witnesses and computes the monitor-coverage curve, validated
against an exhaustive oracle.

Three studies test whether the gap matters in practice. Nine language models, screening
each action of the synthesized witness, approve it almost without exception, yet the
same models flag it when shown both executions. A live open-weights agent commits a
consent-composition CAF in $22$ of $60$ rollouts, which an intent-only guard does not
measurably reduce and a composition-aware guard eliminates. An offline audit of Claude
Sonnet 4.6 finds that it resists every class we test. That agent is safe because the
model behaves, not because a monitor caught anything, and no false-alarm-free
single-execution monitor could have caught a leak, whatever the model.

\smallskip\noindent\textbf{Contributions.}\footnote{All code, synthesized witnesses, the nine-judge sweep data, and reproduction scripts are available at \artifacturl.}
\begin{itemize}\itemsep2pt
\item \emph{A model and a map} (\S\ref{sec:problem}, \S\ref{sec:axes}): CAFs against a
five-conjunct global policy, characterized by expressiveness (safety versus
$k$-safety) and observability (a monitor ladder $P_0$ to $P_7$).
\item \emph{Separation and impossibility} (\S\ref{sec:separation}): the purpose
property is $2$-safety, relational exactly under a stated non-collapse condition, and
invisible to every false-alarm-free per-trace monitor. A relational monitor crosses the
expressiveness barrier but not the statistical one.
\item \emph{A validated synthesizer} (\S\ref{sec:alsynth}, \S\ref{sec:validation}):
\textsc{AL-Synth} computes minimal witnesses and the coverage curve, checked for
soundness, bounded completeness, and minimality against an exhaustive oracle.
\item \emph{Measured monitor blindness} (\S\ref{sec:empirics}): among six surveyed
guardrails, the strongest single-execution profile covers seven of eight classes, six
exactly and explicit labeled flow only conservatively, and misses the relational purpose
class. Per-action judges and a LangGraph guard exhibit the separation, and a live agent shows a composition-aware guard closing a catchable CAF.
\end{itemize}

\noindent\textbf{Relation to classical results.} The ingredients are classical, and we
claim none of them: security properties need not
compose~\cite{mccullough1988noninterference}, noninterference is a
hyperproperty~\cite{goguen1982security,clarkson2010hyperproperties}, execution
monitors enforce only safety~\cite{schneider2000enforceable}, and noninterference can
be enforced precisely by running one execution per security
level~\cite{devriese2010sme}. What is new is what these facts imply for agent
authorization. An agent's authorization policy splits into a trace-safety part and a
relational-purpose part. We give the exact condition under which the purpose part is
relational (Theorem~\ref{thm:2safety}), prove that no false-alarm-free per-trace monitor
detects it, whatever its decision procedure (Theorem~\ref{thm:setlocus}), and extend
this to stochastic agents, for which even cross-execution sampling gives only
statistical detection (Theorems~\ref{thm:distlocus} and~\ref{thm:barrier}). Every
guardrail we survey sits on the trace-safety side. Given a tool schema, a workflow
policy, and a monitor profile, \textsc{AL-Synth} returns a minimal locally permitted
witness outside that monitor's reach and names the observation the monitor would need
to catch it.

%==============================================================================
\section{Threat Model and Background}\label{sec:background}
%==============================================================================
\noindent\textbf{Setting.} We consider a single LLM agent that executes a bounded
episode in response to a user task. The agent issues a finite sequence of actions
(tool calls), each drawn from a typed action set, against an environment with
mutable state. A \emph{local authorization} predicate decides, from the current
state and the action, whether the action is individually permitted. This predicate
abstracts the per-action guards, permission prompts, and tool-call monitors that
deployed agent frameworks use today. A \emph{global policy} $\Phi$ expresses what
the user actually authorized for the episode as a whole.

\noindent\textbf{Adversary.} The adversary's goal is a trace that satisfies every
local check yet violates $\Phi$. We do not assume a jailbroken or
locally-misbehaving agent. The interesting case, and the one our theory targets,
is a benign-looking agent operating under a declared, legitimate purpose whose
\emph{composed behavior} channels information or authority the purpose does not
grant. The adversary may influence the agent through ordinary means such as task
phrasing or injected content in tool results, but it does not need to defeat the
local guards; it needs the composition to do work the guards cannot see.

\noindent\textbf{Running example.} A scheduling assistant is asked to ``find a
time for a team meeting next week'' and is authorized to ``share my free/busy
availability.'' The purpose authorizes disclosure to depend on \emph{whether}
each slot is free. It does not authorize disclosure to depend on \emph{why} a
slot is busy, for instance that a busy block is a confidential interview rather
than an ordinary meeting. Every individual action the assistant takes (read the
calendar, propose slots, send the proposal) can pass a local check. Whether the
\emph{pattern} of what it offers depends on the unauthorized reason is a property
of how it behaves across situations, not of any one message. We return to this example throughout.

\noindent\textbf{Background.} A \emph{trace property} is a set of traces; a
\emph{hyperproperty}~\cite{clarkson2010hyperproperties} is a set of sets of
traces, and a $k$-\emph{safety} hyperproperty has a counterexample consisting of
$k$ finite traces. Noninterference~\cite{goguen1982security} is the canonical
$2$-safety property: it forbids high inputs from affecting low outputs, and a
violation is exhibited by two runs that agree on low inputs but differ on low
outputs. Realistic policies relax strict noninterference through
\emph{declassification}~\cite{sabelfeld2009declassification}, which specifies what high information may be released. An \emph{execution monitor} observes a single run and may halt it, so it can enforce only safety properties of single traces~\cite{schneider2000enforceable,hamlen2006computability}. 

%==============================================================================
\section{Compositional Authorization Failures}\label{sec:problem}
%==============================================================================
\noindent\textbf{Model.} An agent model is a typed transition system. A state
assigns values to a finite set of typed variables. An action $a$ carries a guard
$g_a$, a predicate on states, and an effect that updates a subset of variables.
A finite trace $t=a_1 a_2 \cdots a_n$ is \emph{locally authorized} if, executing
from the initial state, each $a_i$ is enabled (its guard holds) in the state it is
taken from. We write $\Sigma^{*}$ for finite traces. For a trace we distinguish
three projections: $\Low(t)$, the declared purpose and other low inputs; $\Hi(t)$,
the sensitive context the agent may read; and $\Obs(t)$, the outward disclosure
the agent produces.

\begin{definition}[Compositional authorization failure]\label{def:caf}
Fix a model with local authorization predicate $\mathit{auth}$ and global policy
$\Phi$. A \emph{CAF witness} is a finite, locally authorized object that witnesses a
violation of $\Phi$: when $\Phi$ is a trace property, a single trace $t$ with
$\mathit{auth}$ true at every step yet $t \not\models \Phi$; when $\Phi$ is a
hyperproperty, a finite tuple or set of traces, each locally authorized at every step,
that jointly violate $\Phi$. We call the individual traces of a witness its
\emph{executions}. For the relational purpose class the witness is a pair of executions
that are each compliant in isolation.
\end{definition}

\noindent\textbf{The global policy.} We take $\Phi$ to be a conjunction of five
authorization concerns drawn from the access-control and privacy literature:
\[
\Phi \;=\; C_1 \wedge \Ctwo \wedge C_3 \wedge C_4 \wedge C_5,
\]
namely intent ($C_1$: actions stay within the user's task), purpose ($\Ctwo$:
disclosure depends only on purpose-authorized context), provenance ($C_3$: outputs
do not derive from untrusted or cross-domain sources in disallowed ways), consent
($C_4$: disclosure stays within consented scope, per action and cumulatively),
and delegation ($C_5$: delegated authority is exercised only when revalidated).
Four of the five are properties of a single trace. The purpose conjunct is not.

\noindent\textbf{The purpose conjunct.} Following the declassification
view~\cite{sabelfeld2009declassification}, we parameterize purpose by a
\emph{declassification} function $D$, the purpose-authorized view of context, and
an \emph{abstraction} $\alpha$, the policy-relevant view of disclosure:
\begin{equation}\label{eq:c2}
\begin{aligned}
\Ctwo \;=\; \big\{\, S : \forall\, t,t'\!\in S.\;\;
   & \Low(t)=\Low(t')\\[-1pt]
   &{}\wedge\, D(\Hi(t))=D(\Hi(t'))\\[-1pt]
   &{}\Rightarrow\; \alpha(\Obs(t))=\alpha(\Obs(t')) \,\big\}.
\end{aligned}
\end{equation}
In words: two runs that agree on the low inputs and on everything the purpose
authorizes about context must agree on what they disclose. If they can differ,
disclosure is depending on something the purpose did not authorize.

\noindent\textbf{The pair $(D,\alpha)$ comes from the policy.} In the running example,
the purpose ``schedule a team meeting'' with consent ``share free/busy availability''
fixes $D$ as the free/busy status of each slot, excluding the reasons slots are busy.
The abstraction $\alpha$ keeps the policy-relevant parts of a disclosure: which slots
are offered, any attached reason or label, the granularity, and the recipient.
Eq.~\eqref{eq:c2} then says that, with the free/busy view fixed, the offered slots must
not depend on the reasons. For realistic workflows, specifying $(D,\alpha)$ can take
substantial human judgment. The impossibility does not depend on getting it right:
Theorem~\ref{thm:setlocus} holds for every pair, and violations exist exactly when the
pair is residual-nondegenerate (Definition~\ref{def:rdf}, Theorem~\ref{thm:2safety}). A
misspecified pair changes which disclosures count as violations, and so what a
relational auditor flags, but no choice of $(D,\alpha)$ makes a nontrivial purpose
conjunct visible to a false-alarm-free per-action guard.

%==============================================================================
\section{A Two-Axis Characterization}\label{sec:axes}
%==============================================================================
The failure classes differ along two independent axes.

\noindent\textbf{Axis 1: expressiveness.} A class is either a property of a single
trace (a safety property, with a finite bad prefix as a witness) or a relational
property (a $k$-safety hyperproperty, with a $k$-trace witness). Intent,
provenance-by-path, consent, and delegation are single-trace. The purpose
conjunct $\Ctwo$ is relational, and we prove this in \S\ref{sec:separation}.

\noindent\textbf{Axis 2: observability.} A monitor can only act on the state it is
instrumented to observe. We arrange instrumentation as a ladder of profiles
$P_0$ through $P_7$, each strictly more observant than the last. $P_0$ is pure
black box. Intermediate profiles add, in turn, an approximation of user intent,
information-flow labels and a purpose label, a provenance graph, per-action
consent state, a delegated-authority and revalidation view, and cumulative
digests of intent and consent. $P_6$ adds a cross-domain state-transfer view, the
strongest \emph{single-execution} instrumentation we consider. $P_7$ is the only
profile with \emph{relational} power: it can compare distinct executions of the
agent. A monitor at profile $P$ \emph{detects} a class if it can both express the
class (Axis 1) and observe the state the class depends on (Axis 2).

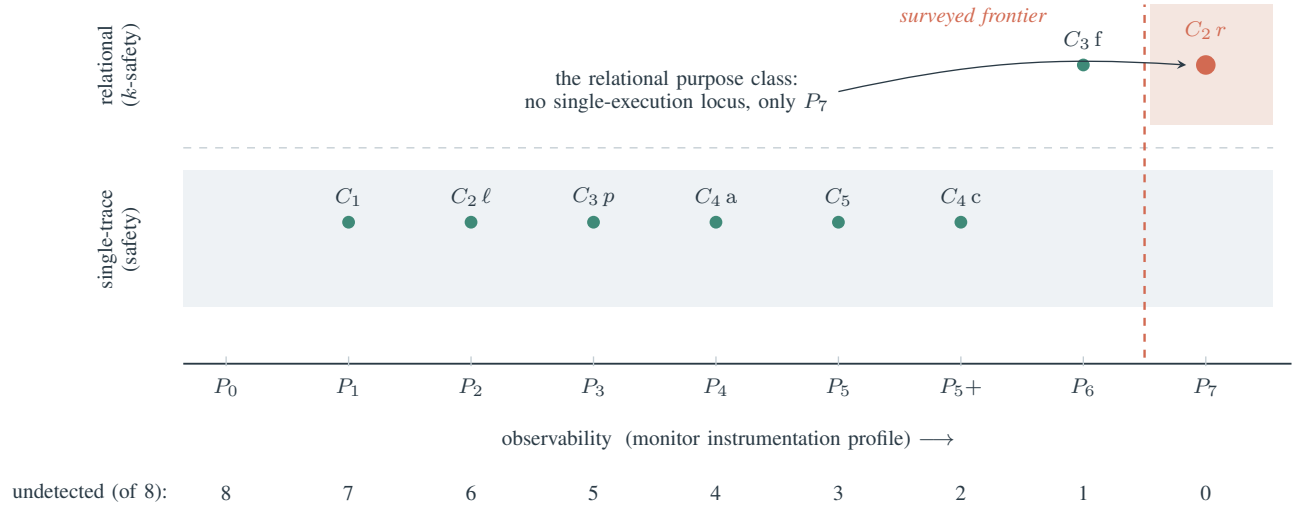
\begin{figure*}[t]
\centering
\begin{tikzpicture}[font=\footnotesize,x=1.62cm,y=1.60cm]
  % bands
  \fill[mist] (-0.35,-0.08) rectangle (8.55,1.05);
  \fill[paleC] (7.55,1.42) rectangle (8.55,2.42);
  \draw[ruleg,dashed,line width=0.5pt] (-0.35,1.235) -- (8.55,1.235);
  % axes
  \draw[ink,line width=0.7pt] (-0.35,-0.55) -- (8.7,-0.55);
  % profile ticks + labels
  \foreach \x/\lab in {0/$P_0$,1/$P_1$,2/$P_2$,3/$P_3$,4/$P_4$,5/$P_5$,6/{$P_5{+}$},7/$P_6$,8/$P_7$}{
     \draw[ruleg,line width=0.5pt] (\x,-0.57) -- (\x,-0.49);
     \node[ink,below=1pt,font=\footnotesize] at (\x,-0.57) {\lab};}
  \node[ink,font=\footnotesize] at (4.1,-1.18) {observability \;(monitor instrumentation profile) $\longrightarrow$};
  % lane labels (left)
  \node[ink,rotate=90,anchor=south,font=\footnotesize,align=center] at (-0.62,0.48) {single-trace\\(safety)};
  \node[ink,rotate=90,anchor=south,font=\footnotesize,align=center] at (-0.62,1.92) {relational\\($k$-safety)};
  % surveyed frontier
  \draw[coral,dashed,line width=0.9pt] (7.5,-0.5) -- (7.5,2.42);
  \node[coral,font=\footnotesize\itshape,anchor=south] at (6.1,2.16) {surveyed frontier};
  % dots
  \tikzset{sdot/.style={circle,fill=teal,inner sep=1.7pt},
           rdot/.style={circle,fill=coral,inner sep=2.6pt}}
  \node[sdot,label={[ink,font=\footnotesize]above:{$C_2\,\ell$}}] at (2,0.62){};
  \node[sdot,label={[ink,font=\footnotesize]above:{$C_3\,p$}}] at (3,0.62){};
  \node[sdot,label={[ink,font=\footnotesize]above:{$C_4\,\mathrm{a}$}}] at (4,0.62){};
  \node[sdot,label={[ink,font=\footnotesize]above:{$C_5$}}] at (5,0.62){};
  \node[sdot,label={[ink,font=\footnotesize]above:{$C_1$}}] at (1,0.62){};
  \node[sdot,label={[ink,font=\footnotesize]above:{$C_4\,\mathrm{c}$}}] at (6,0.62){};
  \node[sdot,label={[ink,font=\footnotesize]above:{$C_3\,\mathrm{f}$}}] at (7,1.92){};
  \node[rdot] at (8,1.92){};
  \node[coral,font=\footnotesize,anchor=south] at (8,2.06){$C_2\,r$};
  % blind-spot annotation
  \node[ink,font=\footnotesize,align=center,anchor=east] (bs) at (5.0,1.7)
       {the relational purpose class:\\no single-execution locus, only $P_7$};
  \draw[->,ink,line width=0.5pt] (bs.east) to[bend left=8] (7.84,1.92);
  % undetected curve
  \node[ink,font=\footnotesize,anchor=east] at (-0.4,-1.62) {undetected (of 8):};
  \foreach \x/\u in {0/8,1/7,2/6,3/5,4/4,5/3,6/2,7/1,8/0}{
     \node[ink,font=\footnotesize] at (\x,-1.62) {\u};}
\end{tikzpicture}
\caption{The two-axis map, with class labels as in Table~\ref{tab:defense}. Increasing observability (left to right) covers the
single-trace classes one by one (teal). The relational purpose class
$C_2\,r$ (coral) is $2$-safety, so no single-execution profile can
express it; it is covered only by the relational profile $P_7$, beyond the
strongest profile any surveyed system reaches. The bottom row is the number of
undetected classes per profile, the monitor-coverage curve of
\S\ref{sec:validation}.}
\label{fig:map}
\end{figure*}

Figure~\ref{fig:map} places the eight subclasses on these axes. The single-trace classes
are covered one by one as instrumentation grows, and explicit labeled flow is covered,
conservatively, at $P_6$. The relational purpose class is $2$-safety, so no
single-execution profile can express it. It is covered only at $P_7$, beyond every
system we survey.

%==============================================================================
\section{Separation Results}\label{sec:separation}
%==============================================================================
We use the standard hyperproperty machinery~\cite{clarkson2010hyperproperties}.
For a set of traces $T$, write $[T]=\{S : S\subseteq T\}$ for the hyperproperty it
induces; a hyperproperty is \emph{trace-property-induced} if it equals $[T]$ for
some $T$. We sketch each argument here. \ifextended The full proofs are in Appendix~\ref{app:proofs}.\else The full proofs are in the extended version of this paper, in our artifact repository at \artifacturl.\fi

\begin{lemma}[The monitorable region]\label{lem:safety}
$C_1$, the path form of $C_3$, $C_4$, and $C_5$ are safety properties of single traces.
\end{lemma}
Each has the form ``$\forall t\in S.\,\phi(t)$'' for a per-trace predicate $\phi$,
hence equals $[\{t:\phi(t)\}]$, and each violation has a finite bad prefix. These
classes are exactly the ones a per-trace monitor can enforce. The labeled-flow form of
$C_3$, written $C_3^{\mathrm{lf}}$, is itself relational ($2$-safety), but its
protected influence is a syntactic data flow from a labeled source to the output, and a
single-execution information-flow profile ($P_6$) over-approximates it soundly through
taint tracking: $P_6$ enforces the single-trace property ``no labeled value reaches a
low output,'' which \emph{conservatively} covers $C_3^{\mathrm{lf}}$ at the usual cost
of some false positives. That is why $C_3^{\mathrm{lf}}$ closes conservatively at $P_6$.
The implicit, control-flow-dependent fragment of noninterference is not
$C_3^{\mathrm{lf}}$ but the relational purpose class $\Ctwo$. Explicit-flow taint tracking misses it, and tracking control dependencies over-approximates it only by rejecting nearly every disclosure made after unauthorized context is read, compliant or not~\cite{costa2025fides}.

The relational purpose class $\Ctwo$ is different. In a $\Ctwo$ witness the agent
discloses from an authorized view that is \emph{identical} in both worlds ($\Low$ and
$D(\Hi)$ equal). No labeled value reaches the output, so taint tracking has nothing to
follow, and the dependence on the unauthorized reason shows only when the two executions
are compared. A single-execution monitor can over-approximate $\Ctwo$ soundly, for
example by blocking every disclosure once unauthorized context is read, but only by
rejecting compliant traces. The results below make this precise: no single-execution
monitor enforces $\Ctwo$ both completely and without false alarms, and cross-execution
comparison ($P_7$) is the first observable that separates compliant from violating
disclosures.

\begin{definition}[Residual-nondegenerate]\label{def:rdf}
A pair $(D,\alpha)$ is \emph{residual-nondegenerate} if there exist finite traces
$a,b$ with $\Low(a)=\Low(b)$, $D(\Hi(a))=D(\Hi(b))$, and
$\alpha(\Obs(a))\neq\alpha(\Obs(b))$.
\end{definition}
In plain terms, the declared declassification does not fix the disclosure, so
unauthorized context has room to move the output. We call this assumption Residual
Disclosure Freedom (RDF-D), and write RDF-D$_A$ when $a,b$ are restricted to the
locally authorized set $A$. RDF-D characterizes $\Ctwo$ (Theorem~\ref{thm:2safety}), and
RDF-D$_A$ makes the global policy $\Phi=[A]\cap\Ctwo$ relational
(Theorem~\ref{thm:phi}).

\begin{theorem}[Exact characterization of $\Ctwo$]\label{thm:2safety}
For every declassification $D$ and abstraction $\alpha$:
(1) $\Ctwo$ is $2$-safety;
(2) $\Ctwo$ is not trace-property-induced if and only if $(D,\alpha)$ is
residual-nondegenerate;
(3) if $(D,\alpha)$ is collapsing, $\Ctwo$ is the trivial hyperproperty,
satisfied by every system.
\end{theorem}
The key observation is the singleton fact: because equality is reflexive, every
single-trace system $\{t\}$ satisfies $\Ctwo$. If $(D,\alpha)$ is nondegenerate, a
violating set contains a pair $a,b$ as in Definition~\ref{def:rdf}, and finiteness
makes $\{a,b\}$ a $2$-trace witness, which gives both $2$-safety and, by the
singleton fact, non-inducedness. If it is collapsing, the implication in
Eq.~\eqref{eq:c2} holds for all admissible pairs, so every system satisfies the
property and it is the trivial, trace-induced hyperproperty. The condition in
part (2) is exactly what the proof needs, and a constant abstraction $\alpha$
shows it cannot be dropped.

\begin{theorem}[$\Phi$ is relational under RDF-D$_A$]\label{thm:phi}
Let $A=\{t: C_1\wedge C_3\wedge C_4\wedge C_5\}$, so $\Phi=[A]\cap \Ctwo$. If
$(D,\alpha)$ is residual-nondegenerate with witnesses in $A$ (RDF-D$_A$), then $\Phi$
is not trace-property-induced.
\end{theorem}
Take RDF-D$_A$ witnesses $a,b\in A$. Each singleton satisfies $\Ctwo$ by reflexivity and
lies in $[A]$, so $\{a\},\{b\}\in\Phi$. If $\Phi=[U]$ for some $U$, then $a,b\in U$,
hence $\{a,b\}\in[U]=\Phi\subseteq\Ctwo$; but $\{a,b\}\notin\Ctwo$ by construction, a
contradiction. If RDF-D$_A$ fails, $\Phi$ collapses to a single-trace property and
per-trace monitoring suffices. RDF-D formalizes the premise of this paper: useful
assistants keep disclosure freedom beyond what their declared purpose fixes, and the
question is whether that freedom carries protected context.

\subsection{The Locus theorems}
The central result says that a relational purpose violation cannot be pinned to any single execution.

\begin{keyresult}
\begin{theorem}[Set Locus]\label{thm:setlocus}
A per-trace monitor that never raises a false alarm for $\Ctwo$ detects no
violation of $\Ctwo$.
\end{theorem}
\smallskip\noindent Every single trace satisfies $\Ctwo$ (the singleton fact), so
a false-alarm-free per-trace monitor cannot flag any trace at all. \emph{Relational purpose violations have no per-trace locus of blame.}
\end{keyresult}

\noindent Contrapositively, a per-trace monitor that catches a genuine purpose violation
must also flag a compliant trace (Corollary~\ref{cor:design}). An LLM at nonzero
temperature is a distribution over traces, so the right property is distributional
independence rather than set equality. Set-$\Ctwo$ over a stochastic agent's support is
stronger than independence, since it also forbids context-independent randomness, which
leaks nothing, so independence is the deployable target.

\begin{theorem}[Distributional Locus]\label{thm:distlocus}
Let the observable $e$ available to a single-execution monitor contain exactly the
per-execution fields $\Low$, $D(\Hi)$ where exposed, and $\Obs$, but not the
nondeclassified context $H$, not the agent's full conditional distribution, and
not cross-context samples. Over the class of all stochastic agents, no monitor
that sees one such $e$ can soundly and completely decide
$Y \perp H \mid \Low, D(\Hi)$, where $Y=\alpha(\Obs)$.
\end{theorem}
For any single observable $e$ of positive probability, one can construct a safe
agent ($Y\perp H$) and a leaky agent ($Y\not\perp H$) that both assign $e$
positive probability. The monitor sees the same $e$ under both: alarming is a
false alarm on the safe agent, and not alarming is a miss on the leaky one. A
single execution cannot certify distributional dependence model-free.

\begin{theorem}[Two-dimensional barrier]\label{thm:barrier}
Under the ladder taxonomy and RDF-D$_A$, no profile $P_{\le 6}$ soundly and completely
enforces $\Phi$, and none soundly detects $\Ctwo$. For the possibilistic property,
the relational profile $P_7$ crosses the barrier, since it can exhibit the
two-trace witness. For the distributional property the barrier has two dimensions,
and $P_7$ crosses only one of them: it crosses the \emph{expressiveness} dimension,
because collecting cross-context samples escapes the per-trace limit, but it does
not cross the \emph{statistical} dimension, because no finite-sample procedure can
soundly and completely decide arbitrary distributional independence over all
stochastic agents.
\end{theorem}
For the possibilistic property, $P_7$ compares two executions and exhibits the witness of
Theorem~\ref{thm:2safety}. For the distributional property, any finite sample is
consistent with two agents, one with $Y\perp H$ and one without, whose sample
distributions are arbitrarily close, so detection is statistical rather than certain.
Logics, model checking, and runtime verification for hyperproperties are established~\cite{clarkson2014temporal,finkbeiner2015algorithms,agrawal2016runtime}, including a probabilistic logic~\cite{abraham2018hyperpctl}, so $P_7$ as a capability is known. Our
claim concerns deployment: no surveyed agent-authorization monitor implements it.

\begin{corollary}[A design rule]\label{cor:design}
Detecting purpose violations admits exactly three responses: tolerate false
positives, audit offline with cross-context samples, or monitor relationally
(compare executions). A per-trace monitor that is both false-alarm-free and useful for
$\Ctwo$ does not exist.
\end{corollary}

\begin{figure}[t]
\centering
\begin{tikzpicture}[font=\footnotesize]
  \tikzset{blk/.style={draw=ink,line width=0.5pt,rounded corners=2pt,fill=mist,
           minimum width=0.9cm,minimum height=0.56cm,inner sep=1.5pt},
           outblk/.style={draw=coral,line width=0.7pt,rounded corners=2pt,fill=paleC,
           minimum width=1.2cm,minimum height=0.56cm,inner sep=1.5pt,text=coral}}
  \node[ink,anchor=east] (la) at (0,0.62) {World $a$};
  \node[blk,right=0.18cm of la] (a1) {read};
  \node[blk,right=0.24cm of a1] (a2) {aggr.};
  \node[outblk,right=0.24cm of a2] (a3) {offer 10};
  \draw[->,ink,line width=0.5pt] (a1)--(a2); \draw[->,ink,line width=0.5pt] (a2)--(a3);
  \node[ink,anchor=east] (lb) at (0,-0.2) {World $b$};
  \node[blk,right=0.18cm of lb] (b1) {read};
  \node[blk,right=0.24cm of b1] (b2) {aggr.};
  \node[outblk,right=0.24cm of b2] (b3) {offer 11};
  \draw[->,ink,line width=0.5pt] (b1)--(b2); \draw[->,ink,line width=0.5pt] (b2)--(b3);
  \node[coral,font=\large] (neq) at ($(a3.east)!0.5!(b3.east)+(0.42,0)$) {$\neq$};
  \node[coral,anchor=west,align=left] at ($(neq.east)+(0.02,0)$) {$\alpha(\Obs)$\\differs};
  \node[teal,anchor=south,align=center] at ($(a2.north)+(0,0.12)$)
       {same authorized view ($\Low$, $D(\Hi)$ equal),\\only the hidden reason differs};
\end{tikzpicture}
\caption{Why purpose is relational. Two executions agree on the low inputs and on
the purpose-authorized view of context (the same slots are free) and differ only
in a hidden, unauthorized reason. Each offers a free slot, so neither trace alone
is faulty, yet the disclosures differ ($\alpha(\Obs_a)\neq\alpha(\Obs_b)$). That
difference is explained only by the unauthorized reason, so the pair is the
$2$-safety witness for $\Ctwo$: it is visible only by comparing executions, which a
relational ($P_7$) monitor does and a per-action monitor cannot.}
\label{fig:rel}
\end{figure}
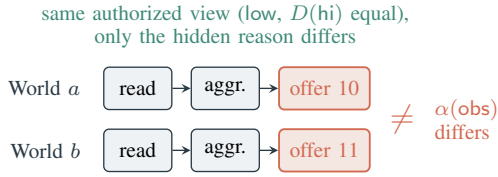

\subsection{An enumeration result, scoped}
Counting, for each profile, the subclasses it cannot detect gives the
\emph{monitor-coverage curve} $[8,7,6,5,4,3,2,1,0]$ from $P_0$ to $P_7$. The endpoints
are proven: $P_0$ detects nothing, $P_7$ covers all eight classes under the
possibilistic semantics of Eq.~\eqref{eq:c2}, and the last class to fall is $\Crel$.
The intermediate values depend on our eight-class taxonomy and ladder. The central
result does not: Theorem~\ref{thm:setlocus} covers every per-trace monitor, whatever it
observes, so under any refinement or coarsening of the single-execution rungs the
relational row stays uncovered until a profile compares executions. Only the number and
order of the intermediate steps change. Stated narrowly: \emph{in our taxonomy, with
delegation scoped to per-action revalidation, exactly one class is not per-trace
monitorable, and it fails by expressiveness, not by any engineering limitation.}

%==============================================================================
\section{\textsc{AL-Synth}: Constructive Synthesis}\label{sec:alsynth}
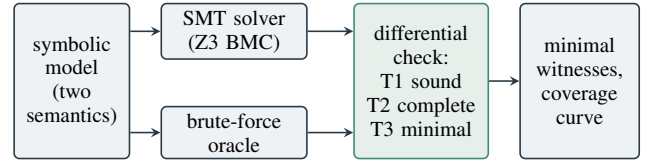
\begin{figure}[t]
\centering
\begin{tikzpicture}[font=\footnotesize,>=stealth,
   bx/.style={draw=ink,line width=0.6pt,rounded corners=2pt,align=center,inner sep=2.5pt,fill=mist},
   ck/.style={bx,fill=paleG,draw=teal}]
  \node[bx,text width=1.35cm,minimum height=2.05cm] (m) {symbolic\\model\\(two\\semantics)};
  \node[bx,text width=1.75cm,anchor=north west] (s) at ($(m.north east)+(0.4,0)$) {SMT solver\\(Z3 BMC)};
  \node[bx,text width=1.75cm,anchor=south west] (o) at ($(m.south east)+(0.4,0)$) {brute-force\\oracle};
  \node[ck,text width=1.6cm,minimum height=2.05cm,anchor=west] (c) at ($(m.east)+(2.95,0)$) {differential\\check:\\T1 sound\\T2 complete\\T3 minimal};
  \node[bx,text width=1.4cm,minimum height=2.05cm,anchor=west] (w) at ($(c.east)+(0.4,0)$) {minimal\\witnesses,\\coverage\\curve};
  \draw[->,ink,line width=0.6pt] (m.east|-s.west) -- (s.west);
  \draw[->,ink,line width=0.6pt] (m.east|-o.west) -- (o.west);
  \draw[->,ink,line width=0.6pt] (s.east) -- (s.east-|c.west);
  \draw[->,ink,line width=0.6pt] (o.east) -- (o.east-|c.west);
  \draw[->,ink,line width=0.6pt] (c.east) -- (w.west);
\end{tikzpicture}
\caption{\textsc{AL-Synth}. One symbolic model with two semantics drives an SMT solver
and an exhaustive oracle. A differential check compares them (soundness, bounded
completeness, minimality), and the validated pipeline emits minimal witnesses and
computes the monitor-coverage curve.}
\label{fig:arch}
\end{figure}

%==============================================================================
\textsc{AL-Synth} makes the gaps concrete. Given a model and a target class, it
synthesizes a minimal CAF witness (a trace, or a pair of traces for the relational
classes) that is locally authorized at every step and globally unauthorized and, for a
target profile, also evades every check that profile can run. The engine is symbolic and
uses no language model. Figure~\ref{fig:arch} shows the pipeline.

\noindent\textbf{Dual semantics.} Each action has two semantics over the same transition
system: a reference semantics, used by an exhaustive oracle, and an SMT encoding, used by
the solver. The harness checks that both induce the same CAFs, so agreement is tested
rather than assumed.

\noindent\textbf{Bounded model checking.} For a horizon $n$, the solver encodes the
initial state, the variable ranges, the transition relation, and the
local-authorization invariant (at every step, the chosen action's guard holds),
then conjoins the global violation at the final state. A satisfying assignment is
a length-$n$ CAF~\cite{biere2003bounded,demoura2008z3}. We enumerate the complete
witness set at a horizon by adding a blocking clause after each model and
re-solving (all-SAT). For synthesis in practice we instead search for the minimal
witness, one satisfiability query per increasing horizon, which is far cheaper.

% fig:arch relocated to Appendix (System Architecture)

\noindent\textbf{Relational witnesses by self-composition.} A relational violation
is existential: it asks for a pair of traces that jointly witness a breach of
$2$-safety. We obtain it by self-composition~\cite{barthe2004selfcomposition}. Two
copies of the system run side by side, and the encoding asserts equality of the
low inputs and of the declassified view, $\Low(a)=\Low(b)$ and
$D(\Hi(a))=D(\Hi(b))$, together with \emph{inequality} of the abstracted
observation, $\alpha(\Obs(a))\neq\alpha(\Obs(b))$. This is exactly the negation of
the $\Ctwo$ implication, so a satisfying assignment is a two-trace witness that the
agent's disclosure depended on context the purpose did not authorize
(Figure~\ref{fig:rel}). Generating a witness is existential, hence SAT. QBF arises only
in the universal direction, proving that a monitor cannot catch a class.

\noindent\textbf{Instrumentation relativization.} To synthesize a witness that is
invisible to a monitor at profile $P$, the encoding conjoins, inside the same
query, the target violation with the negation of every safety check that $P$ can
both express and observe. The evasion is part of the constraint, not a filter
applied afterward.

%==============================================================================
\section{Validation and the Coverage Curve}\label{sec:validation}
%==============================================================================
We validate the solver against the exhaustive oracle on a family of small models,
small precisely so that brute-force enumeration is ground truth.

\noindent\textbf{Differential testing.} Three checks relate the solver to the
oracle, per class and horizon. T1, soundness: every solver witness, replayed
through the reference semantics, is locally authorized at every step and globally
unauthorized. T2, bounded completeness: the solver's all-SAT witness set equals
the oracle's set. T3, minimality: the solver's minimal witness length equals the
oracle's. All three pass for every class, including the relational class, whose
self-composed witness matches the oracle's minimal pair. Oracle agreement is differential testing of two encodings of one intended semantics: it
catches encoding errors, but it does not show that the semantics models a deployment
correctly.

\noindent\textbf{The coverage curve is measured, not assigned.} For each profile and
class, the solver either synthesizes a witness the profile cannot see (a CAF that evades
every check the profile can run or, for the relational classes, the two-trace
counterexample) or certifies that none exists up to the horizon. Solver and oracle agree
on all $9\times 8$ cells (Table~\ref{tab:curve}). Perturbation shows that the curve
tracks the instrumentation: removing the purpose-label observable shifts it to
$[8,7,7,6,5,4,3,2,1]$, as the labeled purpose class staying undetected predicts, and
restoring it returns the original. Against every profile up to $P_6$, the strongest we
survey, the residual is exactly the relational purpose class, an expressiveness gap that
more carried state cannot close.

\begin{table*}[t]
\centering
\caption{Monitor-coverage matrix, solver-validated and confirmed against the
brute-force oracle (all $72$ cells agree). \Y: the profile detects the class exactly,
flagging every witness and no compliant trace. \Yc: detects it only conservatively, a
sound over-approximation that admits most compliant traces at the cost of some false
alarms. \N: no monitor at that profile detects it without missing witnesses or blocking
compliant behavior. The one \Yc\ row is explicit labeled flow, over-approximated by
taint tracking at $P_6$ and above. Both \Y\ and \Yc\ count as covered in the bottom-row curve. Only relational purpose ($C_2$, shaded \Yr) admits no per-trace coverage at all,
closing at $P_7$ (Theorem~\ref{thm:setlocus}).}
\label{tab:curve}
\small
\renewcommand{\arraystretch}{1.25}
\begin{tabular}{lccccccccc}
\toprule
CAF (sub)class & $P_0$ & $P_1$ & $P_2$ & $P_3$ & $P_4$ & $P_5$ & $P_5{+}$ & $P_6$ & $P_7$\\
\midrule
$C_1$ intent                  & \N & \Y & \Y & \Y & \Y & \Y & \Y & \Y & \Y\\
$C_2$ purpose (labeled)       & \N & \N & \Y & \Y & \Y & \Y & \Y & \Y & \Y\\
\rowcolor{paleC!60}$C_2$ purpose (relational)    & \N & \N & \N & \N & \N & \N & \N & \N & \Yr\\
$C_3$ provenance (path)       & \N & \N & \N & \Y & \Y & \Y & \Y & \Y & \Y\\
$C_3$ flow (explicit labeled)  & \N & \N & \N & \N & \N & \N & \N & \Yc & \Yc\\
$C_4$ consent (per-action)    & \N & \N & \N & \N & \Y & \Y & \Y & \Y & \Y\\
$C_4$ consent (composition)   & \N & \N & \N & \N & \N & \N & \Y & \Y & \Y\\
$C_5$ delegation              & \N & \N & \N & \N & \N & \Y & \Y & \Y & \Y\\
\midrule
\textbf{undetected (of 8)} & \textbf{8} & \textbf{7} & \textbf{6} & \textbf{5} & \textbf{4} & \textbf{3} & \textbf{2} & \textbf{1} & \textbf{0}\\
\bottomrule
\end{tabular}
\end{table*}

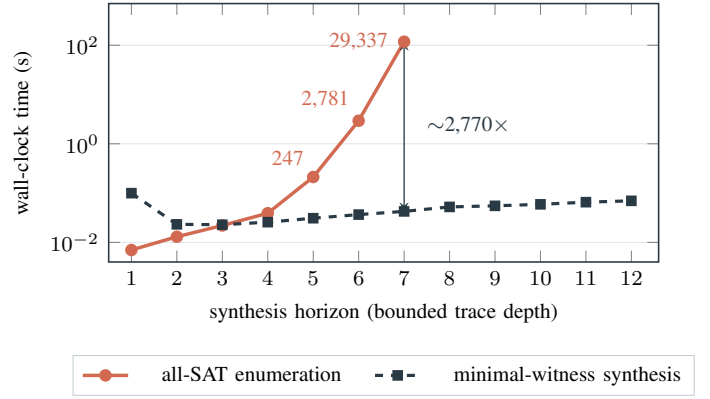
\begin{figure}[t]
\centering
\begin{tikzpicture}
\begin{semilogyaxis}[
  width=\columnwidth, height=\scalingheight,
  xlabel={synthesis horizon (bounded trace depth)}, ylabel={wall-clock time (s)},
  xmin=0.5, xmax=12.6, ymin=0.004, ymax=700, xtick={1,2,3,4,5,6,7,8,9,10,11,12},
  tick label style={font=\footnotesize}, label style={font=\footnotesize},
  ymajorgrids, grid style={ruleg!55,line width=0.3pt}, axis line style={ink,line width=0.6pt},
  legend style={font=\footnotesize, at={(0.5,-0.36)}, anchor=north, legend columns=2,
                draw=ruleg, fill=white, column sep=1em},
  clip=false]
\addplot[coral,line width=1.3pt,mark=*,mark size=1.7pt,mark options={fill=coral,draw=coral}]
  coordinates {(1,0.007)(2,0.013)(3,0.022)(4,0.039)(5,0.212)(6,2.935)(7,117.842)};
\addlegendentry{all-SAT enumeration}
\addplot[ink,line width=1.2pt,dashed,mark=square*,mark size=1.4pt,mark options={solid,fill=ink,draw=ink}]
  coordinates {(1,0.0996)(2,0.0232)(3,0.0228)(4,0.0257)(5,0.0309)(6,0.0366)(7,0.0425)
  (8,0.0524)(9,0.0551)(10,0.0591)(11,0.0654)(12,0.0696)};
\addlegendentry{minimal-witness synthesis}
\node[coral,font=\footnotesize,anchor=south east] at (axis cs:5,0.24){247};
\node[coral,font=\footnotesize,anchor=south east] at (axis cs:6,3.4){2{,}781};
\node[coral,font=\footnotesize,anchor=east] at (axis cs:6.85,117.842){29{,}337};
\draw[<->,ink,line width=0.5pt] (axis cs:7,117.842) -- (axis cs:7,0.0425);
\node[ink,font=\footnotesize,anchor=west] at (axis cs:7.3,2.4) {$\sim$2{,}770$\times$};
\end{semilogyaxis}
\end{tikzpicture}
\caption{Synthesis cost on the cumulative-consent model (11 actions, Z3 4.16).
Minimal-witness synthesis (dashed) stays flat to depth twelve, while all-SAT
enumeration (solid, used only for the oracle check) grows exponentially with the
witness count (annotated). One model; times are illustrative, not a benchmark.}
\label{fig:scaling}
\end{figure}

\noindent\textbf{Cost.} On the cumulative-consent model, minimal-witness synthesis (one
satisfiability query per horizon) stays flat, at about $0.02$ to $0.07$\,s, up to depth
twelve. All-SAT enumeration, used only for the completeness check, grows exponentially:
from $1$ to $29{,}337$ witnesses over depths three to seven, and from $0.02$ to $118$\,s,
a gap of about $2{,}770$ times at depth seven. Cost tracks the query, not the size of
the witness set. This is a shape on one model, not a benchmark. Figure~\ref{fig:scaling} shows both curves, and the data are in the artifact.

% fig:scaling relocated to Appendix (Synthesis Scaling)

%==============================================================================
\section{Empirical Study: Monitor Blindness and Frontier Agents}\label{sec:empirics}
%==============================================================================
We ask two questions. Do agent guardrails detect CAFs, and which ones? And does a
frontier agent commit the relational purpose violation when an adversary sets it up? The
first concerns the monitor, which is what the impossibility is about. The second
concerns the model.

\subsection{The defense matrix: coverage measured per profile}
We do not run the guardrail implementations. We place each surveyed mechanism at the
strongest profile its published design supports and measure that profile with the
campaign of \S\ref{sec:validation}. Table~\ref{tab:survey} lists what each design observes and the profile it receives. None compares two executions of the same purpose\ifextended{} (Appendix~\ref{app:survey})\fi.

\begin{table}[t]
\caption{Surveyed agent-authorization mechanisms and the profile each receives in
Table~\ref{tab:defense}. ``Cross-exec.''\ asks whether the mechanism compares distinct
executions of the same purpose, the $P_7$ capability.}
\label{tab:survey}
\centering
\footnotesize
\renewcommand{\arraystretch}{1.15}
\begin{tabular}{@{}l p{3.0cm} c c@{}}
\toprule
Mechanism & What it observes & Profile & Cross-exec.\\
\midrule
Per-tool approval & current action, shown to a human & $P_0$ & no\\
Task Shield~\cite{jia2025taskshield} & action versus task goal & $P_1$ & no\\
RTBAS~\cite{rtbas2025} & IFC taint and dependencies & $P_3$ & no\\
CaMeL~\cite{debenedetti2025camel} & data and control flow, capabilities & $P_5$ & no\\
Fides~\cite{costa2025fides} & IFC labels, explicit secrecy & $P_6$ & no\\
SAFEFLOW~\cite{safeflow2025} & IFC labels and provenance & $P_6$ & no\\
\midrule
\textsc{AL-Synth} (relational) & two executions, same purpose & $P_7$ & yes\\
\bottomrule
\end{tabular}
\end{table} The assignment is deliberately charitable: crediting a design with less would only widen the gap we report. Each row is therefore a capability bound read from the
design, not a test of the code, and the bound is one-sided where it matters. Running an
implementation could lower a single-trace cell, through a bug for instance, but by
Theorem~\ref{thm:setlocus} it could raise the relational purpose cell only by comparing
executions or by accepting false alarms (Corollary~\ref{cor:design}).

Table~\ref{tab:defense} is read by columns. Coverage grows with sophistication: per-tool
approval catches none of the eight classes, RTBAS three, CaMeL five, and Fides and
SAFEFLOW seven, six exactly and explicit labeled flow conservatively. Yet all six miss
the same class. The relational purpose column is \N\ for every surveyed mechanism,
including Fides, whose authors note that it does not prevent the implicit control-flow
channel this column captures (\S\ref{sec:related}). Only the relational auditor at $P_7$
catches it, as the Locus theorem predicts, whatever model the guard protects.

The witness is minimal. On the scheduling model the shared trace is (collect, aggregate,
report). In world $a$ a busy slot is an ordinary meeting and the agent offers slot $10$.
In world $b$ the same slot is a confidential interview and the agent offers slot $11$.
Each trace alone is locally authorized, so no per-action check fires. The auditor
compares the pair and fires.

\begin{table*}[t]
\centering
\caption{Defense matrix, confirmed against the brute-force oracle; symbols as in
Table~\ref{tab:curve}. Each mechanism is placed at the strongest profile its published
design implements, so its row is a capability bound from the design, not an execution of
its code. The relational purpose column (shaded) is missed by every surveyed mechanism:
Fides and SAFEFLOW ($P_6$) cover six classes exactly and explicit labeled flow
conservatively, and only the relational auditor ($P_7$) covers relational purpose.}
\label{tab:defense}
\small
\renewcommand{\arraystretch}{1.25}
\begin{tabular}{lcccccccc}
\toprule
mechanism (profile) & $C_1$ & $C_2\,\ell$ & \cellcolor{paleC!70}\textbf{$C_2\,r$} & $C_3\,p$ & $C_3\,\mathrm{f}$ & $C_4\,\mathrm{a}$ & $C_4\,\mathrm{c}$ & $C_5$\\
\midrule
Per-tool approval ($P_0$)                       & \N & \N & \cellcolor{paleC!70}\N & \N & \N & \N & \N & \N\\
Task Shield~\cite{jia2025taskshield} ($P_1$)    & \Y & \N & \cellcolor{paleC!70}\N & \N & \N & \N & \N & \N\\
RTBAS~\cite{rtbas2025} ($P_3$)         & \Y & \Y & \cellcolor{paleC!70}\N & \Y & \N & \N & \N & \N\\
CaMeL~\cite{debenedetti2025camel} ($P_5$)       & \Y & \Y & \cellcolor{paleC!70}\N & \Y & \N & \Y & \N & \Y\\
Fides~\cite{costa2025fides} ($P_6$)             & \Y & \Y & \cellcolor{paleC!70}\N & \Y & \Yc & \Y & \Y & \Y\\
SAFEFLOW~\cite{safeflow2025} ($P_6$)      & \Y & \Y & \cellcolor{paleC!70}\N & \Y & \Yc & \Y & \Y & \Y\\
\midrule
\textsc{AL-Synth} relational audit ($P_7$)      & \Y & \Y & \cellcolor{paleC!70}\Yr & \Y & \Yc & \Y & \Y & \Y\\
\bottomrule
\end{tabular}
\end{table*}

\subsection{Case study: the blind spot in a LangGraph workflow}
To check that the blind spot is not an artifact of our abstract model, we reproduce it in
a scheduling workflow built on LangGraph~\cite{langgraph2026}, a widely used agent
framework. The graph (read, aggregate, propose) is gated by a per-action guard that runs
the standard tool-permission check plus purpose-label, provenance, and per-action consent
checks. Two worlds share the authorized free/busy view and differ only in the hidden
reason a slot is busy. The guard returns the same decisions (allow, allow, allow) in both
worlds, although the action arguments differ, and the agent offers slot $10$ in one world
and slot $11$ in the other. The \textsc{AL-Synth} relational auditor compares the runs
and flags the $\Ctwo$ violation. We claim no vulnerability in LangGraph: the blindness
belongs to the per-action permission pattern, and the dependence could be a benign
tie-break, an adversarial choice, or an injected instruction, which no per-action guard
can tell apart.

\subsection{A frontier agent, audited offline}
The theorem says a per-trace monitor cannot catch the relational class. It does not say
whether real agents commit it. We test this directly, acting as an offline auditor with
the cross-context samples a deployed monitor lacks, on Claude Sonnet 4.6 used as a
tool-calling agent at fixed temperature (Table~\ref{tab:claude}). For intent, an
injected instruction in a tool result asks the agent to forward output to an unintended
recipient, and it refuses every time. For consent, the agent stays within a narrow
authorized window every time, and on a deliberately vague variant it asks for
clarification instead of calling a tool. For relational purpose, a paired, interleaved,
pre-registered experiment holds the free/busy view fixed across two conditions that
differ only in the hidden reason, and gates on a high share rate so that a null reflects
disclosure rather than refusal. The agent does not leak.

\begin{table}[t]
\centering
\caption{Offline audit of Claude Sonnet 4.6 (fixed temperature). CAF is the number
of compositional authorization failures observed; UB is the exact one-sided $95\%$
Clopper--Pearson upper bound on the per-episode rate. Prompts, seeds, and settings
are in the artifact.}
\label{tab:claude}
\small
\renewcommand{\arraystretch}{1.2}
\begin{tabular}{llccc}
\toprule
Task & Class & $n$ & CAF & $95\%$ UB\\
\midrule
Injected-doc forward & $C_1$ & $10$ & $0$ & $\le 26\%$\\
Narrow-scope share   & $C_4$ & $10$ & $0$ & $\le 26\%$\\
Scheduling audit     & $C_2\,r$ & $80$ & $0^{\dagger}$ & $\le 3.7\%$\\
\bottomrule
\end{tabular}
\\[3pt]
{\footnotesize $^{\dagger}$Under the policy-relevant abstraction $\alpha$. An initial
parser-artifact positive was removed (see text).}
\end{table}

\paragraph{A false positive, kept on purpose} An initial analysis reported a
significant covert signal, stable across seeds. Raw-output inspection overturned it:
both conditions offered the identical set of slots, and the signal came from a feature
extractor counting a word whose phrasing happened to differ. Under the policy-relevant
abstraction $\alpha$ that Eq.~\eqref{eq:c2} requires, the channel is clean. We keep the
episode because a result is a \emph{security} result only under an abstraction that
tracks the policy, with raw-data inspection as a check that no statistical rigor
replaces.

\subsection{Per-action judges approve the witness; the paired auditor catches it}
The defense matrix places mechanisms by design. Here we put the witness in front of real
models. Nine language models judge the synthesized relational witness step by step:
eight hosted models (from Anthropic, OpenAI, Google, xAI, DeepSeek, Alibaba, Meta, and
Moonshot) and one local open-weights model. A bank of ten scenarios varies the slot pair,
the hidden reason, the task phrasing, and the A/B order, at three repetitions each, which
measure stability rather than add coverage. All judges run at temperature $0$ with a
$4000$-token budget, the hosted ones through OpenRouter's default routing with no pinned
provider. Exact model identifiers are listed under Table~\ref{tab:judges}.

Eight judges never block the witness ($0/30$ each). Kimi-K2.6 blocks it in $3$ of $180$
individual steps ($3$ of $30$ scenario repetitions, across $3$ of the $10$ templates).
The discriminating control is a benign trace in which both worlds disclose the same slot.
DeepSeek blocks $2$ of its $180$ benign steps and Kimi $1$, so Kimi's witness blocks sit
in the same band as its benign blocks. They reflect single-step conservatism, not
relational detection, since a per-action judge never sees the second execution. The
judges are not merely permissive: all nine block an overt single-trace $C_1$ step, an
injected external forward, $30/30$ (Table~\ref{tab:judges}). The relational auditor,
the same model shown both executions, recovers detection: $30/30$ for eight judges and
$17/30$ for Llama-4-Maverick. The table therefore instantiates the observation-model
separation, the same model blind per action and sighted on the pair, rather than nine
independent vulnerabilities, and by the Locus theorem the per-action blindness holds for
any single-execution decision procedure. Llama's $17/30$ also separates relational
\emph{observability}, which $P_7$ supplies, from relational \emph{detection
reliability}, which remains a question of model capability.

\begin{table}[t]
\centering
\caption{Per-action judges over the synthesized relational witness. Cells count the
scenario-repetitions, out of $30$ (ten templates, three repetitions), in which the judge
blocks at least one action: \textbf{wit} the witness, \textbf{ben} a benign no-leak
control, \textbf{C1} an overt single-trace control, and \textbf{aud} the relational
auditor, the same model shown both executions. Hosted models ran through OpenRouter and
the local model through vLLM.}
\label{tab:judges}
\small
\renewcommand{\arraystretch}{1.2}
\begin{tabular}{lcccc}
\toprule
judge & wit & ben & C1 & aud\\
\midrule
Claude Opus 4.8    & $0/30$ & $0/30$ & $30/30$ & $30/30$\\
GPT-5.5            & $0/30$ & $0/30$ & $30/30$ & $30/30$\\
Gemini 3.1 Pro     & $0/30$ & $0/30$ & $30/30$ & $30/30$\\
Grok 4.3           & $0/30$ & $0/30$ & $30/30$ & $30/30$\\
DeepSeek V4 Pro    & $0/30$ & $2/30$ & $30/30$ & $30/30$\\
Qwen3.7-max        & $0/30$ & $0/30$ & $30/30$ & $30/30$\\
Llama-4-Maverick   & $0/30$ & $0/30$ & $30/30$ & $17/30$\\
Kimi-K2.6          & $3/30$ & $1/30$ & $30/30$ & $30/30$\\
Qwen3-30B (local)  & $0/30$ & $0/30$ & $30/30$ & $30/30$\\
\bottomrule
\end{tabular}
\\[4pt]
{\footnotesize\setlength{\tabcolsep}{4pt}%
\begin{tabular}{@{}ll@{}}
\toprule
judge & API identifier (accessed June 2026)\\
\midrule
Claude Opus 4.8    & \texttt{anthropic/claude-opus-4.8}\\
GPT-5.5            & \texttt{openai/gpt-5.5}\\
Gemini 3.1 Pro     & \texttt{google/gemini-3.1-pro-preview}\\
Grok 4.3           & \texttt{x-ai/grok-4.3}\\
DeepSeek V4 Pro    & \texttt{deepseek/deepseek-v4-pro}\\
Qwen3.7-max        & \texttt{qwen/qwen3.7-max}\\
Llama-4-Maverick   & \texttt{meta-llama/llama-4-maverick}\\
Kimi-K2.6          & \texttt{moonshotai/kimi-k2.6}\\
Qwen3-30B (local)  & \texttt{Qwen/Qwen3-30B}\\
\bottomrule
\end{tabular}}
\end{table}

\subsection{A live agent commits a catchable CAF, and the right monitor closes it}
The judge sweep shows that the relational class is invisible per action. A complementary
question is whether a live agent commits a class that a suitably instrumented guard
\emph{can} catch, and whether the guard that closes it is the composition-aware one
rather than a more cautious per-action guard. We run Qwen3-30B as a tool-calling agent
on the cumulative-consent calendar task at temperature $0.7$, with $60$ independent
rollouts per condition, all locally authorized and free of execution errors, under three
guards: none, an intent-closure guard that screens each action against the declared
purpose, and the composition-aware $P_5{+}$ guard, which denies an action once
cumulative disclosure exceeds the workflow's consent bound.

With no guard, the agent commits the $C_4$ consent-composition CAF in $22$ of $60$
rollouts ($37\%$, Wilson $95\%$ CI $[26,49]$) and never on a clean-intent baseline or a
benign control ($0/60$ each), so the failure is compositional rather than a local-guard
error (Table~\ref{tab:livedefense}). The intent-closure guard leaves $20/60$, $3.3$
points below no guard (Newcombe $95\%$ CI $[-13.4,19.8]$, Fisher exact $p=0.85$), as
expected, since the disclosure is not visible in any single action's purpose. The
composition-aware guard removes every offending trace ($0/60$, $33.3$ points below
intent-closure, CI $[21.1,45.9]$, $p<10^{-6}$) while blocking $0/60$ benign and $0/60$
clean-intent rollouts, so no false positives were observed (one-sided $95\%$ upper bound $4.9\%$ for each control). We read this as the implementation
realizing the profile's intended capability, not as a separate causal finding. A second
agent, Gemma-4, never commits the CAF ($0/60$ throughout): the blind spot is structural,
while its exploitation depends on the model.

\begin{table}[t]
\centering
\caption{Live tool-calling agents on the cumulative-consent task ($60$ independent
rollouts per cell at fixed temperature, zero execution errors). \textbf{CAF}:
consent-composition failures committed. \textbf{CAF-free}: fraction of rollouts with no
CAF, a natural rate under no guard and a prevention rate under a guard. The
composition-aware $P_5{+}$ guard removes every CAF while blocking no benign or
clean-intent rollouts.}
\label{tab:livedefense}
\small
\renewcommand{\arraystretch}{1.2}
\setlength{\tabcolsep}{4pt}
\begin{tabular}{lcccc}
\toprule
 & \multicolumn{2}{c}{Qwen3-30B} & \multicolumn{2}{c}{Gemma-4}\\
\cmidrule(lr){2-3}\cmidrule(lr){4-5}
guard & CAF & CAF-free & CAF & CAF-free\\
\midrule
no guard ($P_0$)             & $22/60$ & $63.3\%$ & $0/60$ & $100\%$\\
intent-closure               & $20/60$ & $66.7\%$ & $0/60$ & $100\%$\\
composition ($P_5{+}$) & $0/60$  & $100\%$  & $0/60$ & $100\%$\\
\bottomrule
\end{tabular}
\end{table}

This CAF belongs to the consent-composition class, which closes at $P_5{+}$, not to the
relational purpose class of the impossibility. For a CAF that a single-execution monitor
can catch, the right monitor is the composition-aware one. For the relational class, no single-execution monitor suffices (Table~\ref{tab:judges}).

\noindent\textbf{What the measurements show together.} The studies answer different
questions and are not one kind of evidence. The judges and the LangGraph guard exhibit
the observation-model separation itself: blind on one execution, sighted on the pair.
The live agent shows a different, catchable class that the matching composition-aware
guard closes. The frontier audit reports how one model behaved on the cases we tested,
a property that no deployed per-action monitor could have certified.

%==============================================================================
\section{Discussion}\label{sec:discussion}
%==============================================================================
\noindent\textbf{Detection requires comparison or false alarms.} Single-trace monitors can
enforce intent, provenance, consent, and delegation checks when sufficiently
instrumented, but not the relational purpose conjunct without false positives
(Corollary~\ref{cor:design}). A designer has three detection options: accept and review
false positives, audit offline with cross-context samples, or monitor relationally by
comparing executions. Each has a cost, and a better single-trace guard is not among the
options. Prevention lies outside the corollary, because it changes what the agent sees
rather than what a monitor checks: generating each disclosure from the
purpose-authorized view alone removes the dependence by construction, the agent analogue
of secure multi-execution~\cite{devriese2010sme}, at the price of an extra execution and
of any utility the unauthorized context would have added.

\noindent\textbf{The model, not the monitor, carries the safety.} In our experiments,
safety comes from the model's behavior. The verifiable component, the monitor, is
provably blind to the relational class, so deployments rely on an unverifiable,
model-specific property that may not survive the next model. This argues for investing
in relational monitoring and offline purpose audits.

\noindent\textbf{Limitations.} Our validation is differential testing on small
models, not a proof of correctness and not a scalability study
(\S\ref{sec:validation}). The symbolic models are clean and the tasks synthetic, which
bounds the empirical claims but not the theorem: the impossibility assumes no model of
any API, and noisier tools only add traces, which leaves the singleton fact intact. The
symbolic model feeds only the synthesizer, and its witness replays on an agent
framework (\S\ref{sec:empirics}). The agent-behavior experiments cover two model
families, the nine-model campaign evaluates judges rather than autonomous tool-using
agents, and the guardrail matrix is design-derived, a bound whose direction
\S\ref{sec:empirics} states. The purpose conjunct is modeled in the declassification
style~\cite{sabelfeld2009declassification} rather than the planning semantics of
purpose~\cite{tschantz2012purpose}, and we view the two as complementary. Measuring the
relational class on realistic multi-tool workloads is the natural next step.

%==============================================================================
\section{Related Work}\label{sec:related}
%==============================================================================
\noindent\textbf{Hyperproperties and their verification.} Clarkson and
Schneider~\cite{clarkson2010hyperproperties} introduced hyperproperties and $k$-safety hyperproperties. Temporal logics for them~\cite{clarkson2014temporal} and model checking~\cite{finkbeiner2015algorithms} followed, with runtime verification of $k$-safety~\cite{agrawal2016runtime} and a logic for probabilistic hyperproperties~\cite{abraham2018hyperpctl}. We use this machinery rather than extend it.

\noindent\textbf{Enforcement and composition.} Schneider's enforceable
policies~\cite{schneider2000enforceable} and the computability classes of enforcement
mechanisms~\cite{hamlen2006computability} delimit what an execution monitor can do. Our
Locus theorems instantiate these limits: agent guardrails are execution monitors, and
relational purpose lies outside their reach. Noncomposability of security
properties~\cite{mccullough1988noninterference}, the access-control safety
problem~\cite{hru1976protection}, and noninterference~\cite{goguen1982security} are the
roots of the questions we ask.

\noindent\textbf{Declassification and purpose.} We parameterize purpose by
declassification in the sense of Sabelfeld and
Sands~\cite{sabelfeld2009declassification}. Tschantz, Datta, and
Wing~\cite{tschantz2012purpose} formalize purpose restrictions through a planning
semantics. We see this as complementary to the information-flow view we adopt, and reconciling the two is open.

\noindent\textbf{LLM-agent security.} Indirect prompt
injection~\cite{greshake2023injection} is the overt, single-trace face of agent
insecurity, and our $C_1$ experiment lives in that space. Recent defenses are
per-execution monitors: task-alignment checking of each
action~\cite{jia2025taskshield}, control- and data-flow isolation with
capabilities~\cite{debenedetti2025camel}, and information-flow taint-tracking
planners~\cite{costa2025fides}. We place these on our ladder in \S\ref{sec:empirics}. None compares distinct executions of the same purpose,
and the strongest of them concedes the gap: Fides~\cite{costa2025fides} enforces
explicit secrecy but, by its authors' own analysis, does not prevent implicit flows
through control-flow decisions, which is exactly the relational purpose channel our
Locus theorem isolates.

%==============================================================================
\section{Conclusion}\label{sec:conclusion}
%==============================================================================
Local authorization does not compose. We formalized the failure along expressiveness and
observability, and proved that the relational purpose violation has no per-trace locus
of blame, so no false-alarm-free per-action guard can detect it. \textsc{AL-Synth} makes
the result constructive: it computes the coverage curve, validated against an exhaustive
oracle, and names the observation each CAF class requires. Nine judges and a LangGraph workflow exhibit the separation in practice, a live agent shows the composition-aware guard closing a catchable CAF, and an offline audit finds no leak by a frontier model on the cases tested. The takeaway is a design rule, not a patch: to bound relational purpose, or any policy conjunct that is genuinely relational on the permitted executions, a monitor must compare executions or accept false alarms.

%==============================================================================
% MANDATORY CFP SECTIONS.
%==============================================================================
\section*{Ethics Considerations}
\textsc{AL-Synth} is dual-use: it synthesizes authorization-evading traces. The risk is
bounded because it works over an abstract model the analyst supplies and emits abstract
action sequences, not exploits, so using it against a real target already requires a
formal model of that target and its monitors. Its purpose is defensive, and the benefit
outweighs the marginal uplift to an attacker who already holds such a model. The study
used the authors' own API accounts and local deployments, touched no third-party data or
production accounts, and collected no human-subjects data. We release the artifact
(\artifacturl) with documentation framing its defensive use.

\section*{AI Disclosure}
Two uses of generative AI are relevant. As a writing aid, a large language model
assistant helped draft and edit the prose across all sections. It did not originate the
theorems, the synthesizer, or the experimental design, which the authors developed,
implemented, and validated, and the authors verified the correctness and originality of
all content, including every reference and reported number.

Generative AI is also the object of study: \S\ref{sec:empirics} evaluates language
models as tool-using agents and as per-action and relational judges, and reports their
settings\ifextended{} (identifiers in Appendix~\ref{app:repro})\fi. Capable models are
necessary because the separation turns on whether disclosures depend on unauthorized
context, which weaker models do not exercise. We minimized query volume, report full
counts, and release all logs. Hosted runs used the authors' own accounts, and
open-weights runs a single GPU.

%==============================================================================
\ifextended
\appendices
\section{Proofs}\label{app:proofs}

\noindent\textit{Lemma~\ref{lem:safety}.}
Each of $C_1$, the path form of $C_3$, $C_4$, and $C_5$ is of the form $\{S:\forall t\in S.\,\phi(t)\}$ for a
per-trace predicate $\phi$, hence equals $[\{t:\phi(t)\}]$ and is trace-induced.
Each $\phi$ is a safety predicate on finite traces with a finite bad prefix
(an out-of-intent action, an untrusted-provenance write, an out-of-scope or
over-threshold disclosure, an unrevalidated delegated action), so each is a safety
property. The labeled-flow form $C_3^{\mathrm{lf}}$ is $2$-safety and not
trace-induced; it is enforced not directly but through the single-trace taint-flow
property of $P_6$, which soundly over-approximates it (conservative coverage, not
exact). \hfill$\blacksquare$

\smallskip
\noindent\textit{Theorem~\ref{thm:2safety}.}
\emph{Singleton fact.} For any $t$, the pair condition in Eq.~\eqref{eq:c2} ranges
over $t,t'\in\{t\}$, i.e.\ $t'=t$, and equality is reflexive, so $\{t\}\in\Ctwo$.

(1) If $(D,\alpha)$ is collapsing there is no admissible pair on which the
consequent fails, so $\Ctwo=2^{\Sigma^{*}}$, which is vacuously $k$-safety for any
$k$. If $(D,\alpha)$ is nondegenerate, let $S\notin\Ctwo$. Then $S$ contains
$a,b$ with $\Low(a)=\Low(b)$, $D(\Hi(a))=D(\Hi(b))$, and
$\alpha(\Obs(a))\neq\alpha(\Obs(b))$. The pair $M=\{a,b\}$ is finite, $M\subseteq
S$, and every superset of $M$ also violates $\Ctwo$; so $M$ is a $2$-trace
witness and $\Ctwo$ is $2$-safety.

(2,$\Leftarrow$) Suppose $(D,\alpha)$ is nondegenerate yet $\Ctwo=[T]$ for some
$T$. By the singleton fact $\{a\},\{b\}\in\Ctwo=[T]$, so $a,b\in T$, hence
$\{a,b\}\subseteq T$ and $\{a,b\}\in[T]=\Ctwo$. But $\{a,b\}\notin\Ctwo$ by choice
of the witness, a contradiction. So $\Ctwo$ is not trace-induced.

(2,$\Rightarrow$) and (3) Suppose $(D,\alpha)$ is collapsing. Then for every set
$S$ and every $t,t'\in S$ the antecedent of Eq.~\eqref{eq:c2} either fails or the
consequent holds, so $S\in\Ctwo$; thus $\Ctwo=2^{\Sigma^{*}}=[\Sigma^{*}]$, which
is trace-induced. Hence if $\Ctwo$ is not trace-induced, $(D,\alpha)$ is
nondegenerate, and the trivial case is exactly the collapsing one.
\hfill$\blacksquare$

\smallskip
\noindent\textit{Theorem~\ref{thm:phi}.}
$\Phi=[A]\cap\Ctwo$. Suppose $\Phi=[U]$. Under RDF-D$_A$ there are $a,b\in A$ with
$\Low(a)=\Low(b)$, $D(\Hi(a))=D(\Hi(b))$, $\alpha(\Obs(a))\neq\alpha(\Obs(b))$.
Since $\{a\},\{b\}\subseteq A$ and singletons satisfy $\Ctwo$, we have
$\{a\},\{b\}\in\Phi=[U]$, so $a,b\in U$, hence $\{a,b\}\in[U]=\Phi\subseteq\Ctwo$.
But $\{a,b\}\notin\Ctwo$, a contradiction. \hfill$\blacksquare$

\smallskip
\noindent\textit{Theorem~\ref{thm:setlocus}.}
Let $M$ be per-trace with accepted set $[T_M]$ and false-alarm-free for $\Ctwo$.
By the singleton fact every $\{t\}\in\Ctwo$, so $M$ may not alarm on any $t$, i.e.\
$T_M=\Sigma^{*}$ and $M$ accepts every trace. Therefore $M$ flags no trace and
detects no violation. \hfill$\blacksquare$

\smallskip
\noindent\textit{Theorem~\ref{thm:distlocus}.}
Fix an observable value $e=(\Low,D(\Hi),y)$ with $y=\alpha(\Obs)$ of positive
probability. Construct agent $\mathcal{A}_0$ whose output is independent of $H$
given $(\Low,D(\Hi))$ and assigns $e$ positive probability, and agent
$\mathcal{A}_1$ whose output depends on $H$ and also assigns $e$ positive
probability under some value of $H$. Both are consistent with the single observed
$e$. A sound-and-complete monitor must alarm on $\mathcal{A}_1$ and not on
$\mathcal{A}_0$, but it sees the same $e$ in both, so any fixed decision on $e$ is
wrong for one of them. Hence no monitor seeing one such $e$ decides the property
soundly and completely over the class of all stochastic agents. \hfill$\blacksquare$

\smallskip
\noindent\textit{Theorem~\ref{thm:barrier}.}
The first two clauses are Theorems~\ref{thm:phi} and~\ref{thm:setlocus} read on
the ladder: a $P_{\le 6}$ monitor is single-execution, so it cannot express
$\Ctwo$, and by Theorem~\ref{thm:phi} it cannot soundly and completely enforce
$\Phi$. For the possibilistic property, a $P_7$ monitor compares two executions
and exhibits the witness of Theorem~\ref{thm:2safety}, crossing the barrier. For
the distributional property, the expressiveness dimension is crossed because
cross-context sampling represents the comparison a per-trace monitor cannot, while
the statistical dimension is not crossed: by a standard indistinguishability
argument, for any finite sample there are two agents, one with $Y\perp H$ and one
without, whose sample distributions are arbitrarily close, so no finite-sample
test decides independence soundly and completely over all agents. Detection is
therefore statistical, not certain. \hfill$\blacksquare$

\section{Monitor-Profile Survey}\label{app:survey}
We map recent and deployed agent-authorization mechanisms to the profile ladder of
\S\ref{sec:axes}. For each, we record from its primary source what per-execution
state it observes, whether it compares distinct executions of the same purpose, and
the resulting profile. Table~\ref{tab:survey} is the result. Every entry is
single-execution: each system inspects the current run (a tool call, its arguments,
data-flow or taint labels, or the running plan), and none compares two executions
of the same purpose to test whether disclosure depends on unauthorized context.
This substantiates the claim that no surveyed monitor reaches $P_7$.

We assign each design the strongest profile its primary source plausibly supports, a
deliberately charitable upper bound: under-crediting a mechanism would only widen the
gap we report, so generosity makes our conclusion harder to reach, not easier. Task
Shield is $P_1$ because it checks each action against the user's task (per-action
intent). RTBAS reaches $P_3$ because it carries information-flow taint and dependency
labels (provenance), and we charitably credit it with the weaker intent and
purpose-label observables a $P_3$ monitor subsumes. CaMeL reaches $P_5$ because it
extracts data and control flow and enforces capabilities (provenance, per-action
consent, and delegation scope). Fides and SAFEFLOW reach $P_6$ because they add
confidentiality and integrity labels with cross-domain reasoning. In every row the
cross-execution column is no for the same structural reason: each mechanism evaluates a
single execution and never compares two executions under matched low inputs and
declassified context, which is the only way to witness a relational purpose violation.
The assignment is thus generous on every axis except the one that matters, and on that
axis every surveyed design sits at zero.

The most advanced surveyed designs are information-flow systems. Taint tracking is conservative by construction, which places these systems in the false-positive corner of Corollary~\ref{cor:design}, and the strongest states its own limit: Fides enforces \emph{explicit secrecy} but does not prevent implicit flows through control-flow decisions~\cite{costa2025fides}. That residual channel is the
relational purpose violation of the Locus theorem, which \textsc{AL-Synth}'s relational
mode ($P_7$) expresses by comparing same-purpose executions.

\section{Judge Reproducibility Metadata}\label{app:repro}
Table~\ref{tab:repro} records the nine-model judge campaign of \S\ref{sec:empirics} in
the detail needed to reproduce it. Eight models were queried through the OpenRouter
chat-completions endpoint and one open-weights model through a local vLLM server. Every
request fixed the sampling temperature at $0$ and allowed a $4000$-token generation
budget, so that reasoning-heavy models had room to deliberate before emitting a
verdict. Requests used OpenRouter's default routing: we did not pin a serving provider
or disable fallback, so a given slug may have been dispatched to whichever provider
OpenRouter selected at request time, a reproducibility caveat we state explicitly. The
per-judge and per-scenario counts behind Table~\ref{tab:judges}, together with every
raw model response, are released with the artifact.

\begin{table}[htbp]
\caption{Judge models and exact API slugs for the campaign of \S\ref{sec:empirics}; settings in the text, accessed June 2026.}
\label{tab:repro}
\small
\renewcommand{\arraystretch}{1.15}
\centering
\begin{tabular}{ll}
\toprule
Display name & API slug\\
\midrule
Claude Opus 4.8     & \texttt{anthropic/claude-opus-4.8}\\
GPT-5.5             & \texttt{openai/gpt-5.5}\\
Gemini 3.1 Pro      & \texttt{google/gemini-3.1-pro-preview}\\
Grok 4.3            & \texttt{x-ai/grok-4.3}\\
DeepSeek V4 Pro     & \texttt{deepseek/deepseek-v4-pro}\\
Qwen3.7-max         & \texttt{qwen/qwen3.7-max}\\
Llama-4-Maverick    & \texttt{meta-llama/llama-4-maverick}\\
Kimi-K2.6           & \texttt{moonshotai/kimi-k2.6}\\
Qwen3-30B$^{\ast}$  & \texttt{Qwen/Qwen3-30B}\\
\bottomrule
\end{tabular}
\\[3pt]
{\footnotesize $^{\ast}$Served locally via vLLM on a single GPU; all other models
through the OpenRouter chat-completions endpoint.}
\end{table}

\fi
\FloatBarrier

\bibliographystyle{IEEEtran}
\bibliography{refs}

\end{document}